\documentclass[preprint,12pt,numbers,sort&compress]{elsarticle}
\usepackage[T1]{fontenc}
\usepackage[utf8]{inputenc}
\usepackage{lmodern}
\usepackage{amsmath,amssymb,amsthm}
\usepackage{microtype,booktabs,tabularx,array}
\usepackage{tikz,float}
\numberwithin{equation}{section}
\usepackage[hidelinks]{hyperref}
\newtheorem{maintheorem}{Theorem}

\newtheorem{lemma}{Lemma}[section]
\newtheorem{theorem}[lemma]{Theorem}
\newcommand{\MDG}{\textsc{MDG}}
\newcommand{\dotcup}{\mathbin{\dot\cup}}

\begin{document}
\begin{frontmatter}
\title{Exact Tie-Breaking and Tight Degree Bounds\\
for Minimum-Degree Greedy on Interval Graphs}
\author[inst1]{Ruize Zhao\texorpdfstring{\corref{cor1}}{}}
\cortext[cor1]{Corresponding author}
\ead{zhaoruize@stu.xjtu.edu.cn}
\address[inst1]{Xi'an Jiaotong University, No. 28 Xianning West Road, Xi'an, Shaanxi 710049, China}

\begin{abstract}
We study tie-breaking in the minimum-degree greedy algorithm (\MDG) for maximum independent set on interval graphs.
Given an interval representation of an $n$-vertex graph, we compute the largest and smallest output sizes and construct executions attaining them in $O(n^3)$ time and $O(n^2)$ space.
For maximum degree at most four, favorable tie-breaking gives a tight guarantee of $3/4$.
For every maximum-degree bound of at least five, the guarantee drops to $2/3$. Both bounds are tight even on connected unit interval graphs whose output size is independent of tie-breaking.
Together with the known optimality of favorable tie-breaking for maximum degree at most three, these results give a complete classification by degree bound.
\end{abstract}

\end{frontmatter}

\section{Introduction}
\label{sec:introduction}

\label{sec:background}

\MDG\ constructs an independent set by repeatedly selecting a minimum-degree vertex and deleting its closed neighborhood. When several vertices have minimum degree, the choice among them can affect the output size. We ask how well, and how poorly, the algorithm can perform on interval graphs. We also determine the guarantee under favorable tie-breaking for each maximum-degree bound.

We use the following quantities to describe the possible outputs. A \emph{legal execution} follows the minimum-degree rule until the graph is empty. For a legal execution $\sigma$, let $|\sigma|$ be the number of selected vertices. Define
\begin{equation}
B(G)=\max_{\sigma\text{ complete and legal}}|\sigma|,
\qquad
W(G)=\min_{\sigma\text{ complete and legal}}|\sigma|,
\label{eq:execution-values}
\end{equation}
with $B(\varnothing)=W(\varnothing)=0$. Thus $W(G)\le B(G)\le\alpha(G)$, where $\alpha(G)$ is the independence number. We allow tie-breaking to use arbitrary information. The only requirement is that the selected vertex have minimum degree in the current graph.

An interval graph is the intersection graph of closed intervals on the real line. It is a unit interval graph if the intervals can all be chosen to have the same positive length. Given the endpoint order, a maximum independent set can be found in linear time by selecting intervals in order of earliest finish~\cite{GuptaLeeLeung1982}. The intervals selected in this way need not have minimum degree, so this algorithm does not tell us how well \MDG\ can perform.

On general graphs of maximum degree $\Delta\ge3$, \MDG\ guarantees a $3/(\Delta+2)$ fraction of the optimum~\cite{HalldorssonRadhakrishnan1997}. Chaplick et al.~\cite{ChaplickEtAl2024} proved that every execution on an interval graph returns at least $2\alpha(G)/3$ vertices. This bound is tight even on unit interval graphs of maximum degree three. They also observed that favorable tie-breaking is optimal for maximum degree at most three, and asked what guarantee it gives for larger degree bounds.

Bodlaender, Thilikos, and Yamazaki~\cite{BodlaenderThilikosYamazaki1997} denoted the largest and smallest legal output sizes by $\mathrm{max\mbox{-}GR}(G)$ and $\mathrm{min\mbox{-}GR}(G)$. They proved hardness results for recognizing graphs on which \MDG\ attains a prescribed approximation ratio. Krysta, Mari, and Zhi~\cite{KrystaMariZhi2024} call the largest-output problem \textsc{MaxGreedy} and write $\alpha^+(G)=B(G)$. They study advice-based tie-breaking and its computational complexity. Halld\'orsson and Yoshihara~\cite{HalldorssonYoshihara1995} studied secondary choice rules and limitations of favorable tie-breaking on low-degree graphs. Section~4 of Chaplick et al.~\cite{ChaplickEtAl2024} asks about the complexity of favorable tie-breaking on interval and chordal graphs.

\label{sec:contributions}

We address both the exact optimization problem and the guarantee under a degree bound. To state the latter precisely, for a nonnegative integer $d$, define
\begin{equation}
c_d=\inf\left\{
\frac{B(G)}{\alpha(G)}:
G\ne\varnothing\text{ is an interval graph and }\Delta(G)\le d
\right\}.
\label{eq:cd}
\end{equation}

We first compute both extreme output sizes.

\begin{maintheorem}[Exact optimization]
\label{thm:A}
Given a closed interval representation of an $n$-vertex graph $G$, one can compute $B(G)$ and $W(G)$ and construct legal executions attaining both values in $O(n^3)$ time and $O(n^2)$ space in the comparison/RAM model.
\end{maintheorem}

For maximum degree four, we obtain the following tight bound.

\begin{maintheorem}[Degree four]
\label{thm:B}
Every interval graph $G$ with $\Delta(G)\le4$ satisfies
$B(G)\ge\frac34\alpha(G).$
The bound is tight even on connected unit interval graphs with $B(G)=W(G)$. In particular, $c_4=3/4$.
\end{maintheorem}

For larger degree bounds, the guarantee is lower.

\begin{maintheorem}[Degree at least five]
\label{thm:C}
For every integer $d\ge5$, we have $c_d=2/3$. This infimum is approached by connected unit interval graphs of maximum degree five with $B(G)=W(G)$.
\end{maintheorem}

Together with the known result for $d\le3$, these theorems give
\begin{equation}
c_d=
\begin{cases}
1, & 0\le d\le3,\\
3/4, & d=4,\\
2/3, & d\ge5.
\end{cases}
\label{eq:degree-classification}
\end{equation}
The same values hold when the infimum is restricted to unit interval graphs. For comparison, under adversarial tie-breaking the guarantee on unit interval graphs is $1$ for maximum degree at most two and $2/3$ for every larger degree bound~\cite[Theorem~2 and Lemma~7]{ChaplickEtAl2024}.

\subsection{Outline of the proofs}
\label{sec:overview}

The algorithm in Section~\ref{sec:exact-dp} rests on a simple property of the interval representation. After \MDG\ selects an interval, every surviving interval lies strictly to its left or strictly to its right. We can solve these two parts independently and interleave their executions while respecting the minimum-degree rule. This gives a dynamic program with $O(n^2)$ subproblems, each taking $O(n)$ time.

The degree-four bound uses induction. Selecting a simplicial minimum-degree vertex reduces the independence number by exactly one, so induction handles this case directly. If no such vertex exists, we show that the connected graph is a unit interval graph.

Even in a unit interval graph, however, a minimum-degree selection may reduce the independence number by two. In Section~\ref{sec:degree-four}, we account for this loss by keeping virtual vertices at the ends of the remaining graph and proving a stronger inductive statement. A family of connected equality graphs shows that the bound cannot be improved. Section~\ref{sec:degree-five} gives a degree-five family whose ratios approach $2/3$.

\section{Preliminaries}
\label{sec:preliminaries}\label{lem:umbrella-properties}

We fix notation and record the ordering properties used in the proofs. All graphs are finite, simple, undirected, and unweighted. For a graph $G=(V,E)$ and vertex $v$, let $N_G(v)$ and $N_G[v]=N_G(v)\cup\{v\}$ denote its open and closed neighborhoods. Write $d_G(v)=|N_G(v)|$, $\delta(G)=\min_{v\in V}d_G(v)$, and $\Delta(G)=\max_{v\in V}d_G(v)$ when $V\ne\varnothing$.

For $X\subseteq V$, let $G-X=G[V\setminus X]$. Vertices outside $G$ in a deletion set have no effect. We set $\alpha(\varnothing)=0$.

A vertex is \emph{simplicial} if its closed neighborhood induces a clique. We follow the interval- and chordal-graph terminology in Golumbic~\cite{Golumbic2004} and write $G_1\dotcup G_2$ for disjoint union with no edges between the parts.

An interval representation assigns a finite closed interval $I_v=[\ell_v,r_v]$, with $\ell_v\le r_v$, to each vertex. Distinct vertices are adjacent exactly when their intervals intersect. We allow repeated endpoints, identical intervals, and point intervals. In particular, intervals touching at an endpoint are adjacent. Unit interval graphs are also called indifference graphs~\cite{Roberts1969}. Scaling a representation of common positive length to length one preserves the graph. Following the standard orderings of unit interval graphs~\cite{Golumbic2004,LoogesOlariu1993}, call an order an \emph{umbrella order} if
\begin{equation}\label{eq:umbrella}
x<y<z,\quad xz\in E(G)\quad\Longrightarrow\quad xy,yz\in E(G).
\end{equation}
Ordering equal-length intervals by nondecreasing left endpoint, with arbitrary ties, gives an umbrella order, and restricting this order to an induced subgraph preserves the umbrella property. Several useful facts follow. The vertices spanned by an edge form a clique. Each closed neighborhood is consecutive, and the neighbors on either side form a clique. Thus the first and last vertices are simplicial, and an independent set meets any closed neighborhood in at most two vertices.

The components form consecutive blocks in the order. In a connected graph, consecutive vertices are adjacent, so every nonempty prefix or suffix is connected. We use these properties only for unit interval graphs.

\subsection{Legal interleaving and additivity}
\label{sec:interleaving}

\begin{lemma}\label{lem:interleave}\label{lem:restriction}
Suppose $V(G)$ is partitioned into sets $V_1,\dots,V_k$ with no edges between distinct parts, and let $G_i=G[V_i]$. Prescribed legal executions on the $G_i$ can be interleaved, preserving each internal order, to give a legal execution on $G$. Conversely, every legal execution on $G$ restricts to one on each $G_i$. Consequently,
\[
B(G)=\sum_{i=1}^k B(G_i),\qquad W(G)=\sum_{i=1}^k W(G_i).
\]
\end{lemma}

\begin{proof}
We construct an interleaving by comparing the next prescribed vertices in the unfinished parts. Each has minimum degree within its part, so choosing one of smallest degree among them gives a legal selection in $G$. Since a selection in one part has no effect on the others, we can repeat this procedure until all parts are empty.

Conversely, any vertex selected in $G$ has minimum degree in its remaining part. Only selections in that part can delete its vertices, so the restricted execution is a legal execution. Its length lies between $W(G_i)$ and $B(G_i)$. Summing gives the stated bounds, and interleaving optimal executions attains them. The interleaving matters here: simply concatenating executions need not be legal.
\end{proof}

\section{Computing the best and worst output sizes}
\label{sec:exact-dp}

We first prove Theorem~\ref{thm:A}. The algorithm applies to interval graphs of arbitrary degree and does not require equal interval lengths. We assume that an interval representation is given. If necessary, one can recognize an interval graph and construct a representation in $O(n+|E(G)|)$ time~\cite{BoothLueker1976}.

\subsection{The subproblems}

Let $e_1<e_2<\cdots<e_m$ be the distinct endpoint coordinates, where $m\le2n$, and add $e_0=-\infty$ and $e_{m+1}=+\infty$. For each vertex $v$, write $\ell_v=e_{p_v}$ and $r_v=e_{q_v}$, where $p_v\le q_v$. For $0\le i<j\le m+1$, define
\[
S_{ij}=\{v:e_i<\ell_v\text{ and }r_v<e_j\},\qquad
G_{ij}=G[S_{ij}],\qquad n_{ij}=|S_{ij}|.
\]
We call $G_{ij}$ a \emph{window}. Its defining quantities are:
\begin{itemize}
\item $S_{ij}$, the vertices whose intervals lie strictly between the two endpoints;
\item $G_{ij}$, the graph induced by these vertices;
\item $n_{ij}$, the number of these vertices.
\end{itemize}
There are $O(n^2)$ windows, and $G_{0,m+1}=G$. Recall that intervals touching at an endpoint are adjacent, which is why the inequalities are strict. The next lemma shows that selecting a vertex leaves two child windows.

\begin{lemma}\label{lem:window-closure}
If $v\in S_{ij}$, then

\begin{equation}
G_{ij}-N_{G_{ij}}[v]
=G_{i,p_v}\ \dot\cup\ G_{q_v,j},
\label{eq:window-closure}
\end{equation}

where $\dot\cup$ denotes disjoint union with no crossing edges.
\end{lemma}

\begin{proof}
We identify the surviving intervals on each side of $v$. A vertex $u\in S_{ij}$ survives precisely when $r_u<\ell_v$ or $\ell_u>r_v$. The first condition gives $S_{i,p_v}$, and the second gives $S_{q_v,j}$. Both sets lie in the parent window, since $i<p_v\le q_v<j$. There are no edges between them: every interval in the first set ends before every interval in the second set starts.
\end{proof}

The two child windows contain exactly the nonneighbors of $v$ in $S_{ij}\setminus\{v\}$. Thus its current degree is
\begin{equation}
d_{G_{ij}}(v)=n_{ij}-1-n_{i,p_v}-n_{q_v,j}.
\label{eq:window-degree}
\end{equation}\label{lem:degree-count}

For a nonempty window, let $\delta_{ij}$ be its minimum degree and let $M_{ij}$ be the set of vertices of degree $\delta_{ij}$. Thus $M_{ij}$ consists of the possible first choices of \MDG\ on $G_{ij}$.

Set $U_{ij}=B(G_{ij})$ and $L_{ij}=W(G_{ij})$. For an empty window, $U_{ij}=L_{ij}=0$. For nonempty windows, use the recurrences

\begin{equation}
U_{ij}=1+\max_{v\in M_{ij}}
\bigl(U_{i,p_v}+U_{q_v,j}\bigr),
\label{eq:best-recurrence}
\end{equation}

\begin{equation}
L_{ij}=1+\min_{v\in M_{ij}}
\bigl(L_{i,p_v}+L_{q_v,j}\bigr).
\label{eq:worst-recurrence}
\end{equation}

\begin{lemma}\label{lem:recurrences}
Equations \eqref{eq:best-recurrence} and \eqref{eq:worst-recurrence} are exact.
\end{lemma}

\begin{proof}
We induct on $|S_{ij}|$ and condition on the first choice. The empty case is immediate. A legal execution starts at some $v\in M_{ij}$ and leaves the two smaller windows in Lemma~\ref{lem:window-closure}. By Lemma~\ref{lem:interleave}, the best remaining length is the sum of their best lengths, and the worst remaining length is the sum of their worst lengths. Interleaving the corresponding child executions attains both sums. It remains only to take the maximum or minimum over the possible first choices, which gives the recurrences.
\end{proof}

\subsection{Evaluation and reconstruction}

For the time bound, we count endpoint comparisons and operations on integers of $O(\log n)$ bits. No arithmetic on arbitrary-precision coordinates is required. After sorting the endpoints, the counts $n_{ij}$ can be computed in $O(n^2)$ time. To do this, initialize a zero table $A$ and increment $A[p_v-1,q_v+1]$ for each interval $v$. Taking suffix sums in the first coordinate and prefix sums in the second gives
\[
n_{ij}=\sum_{x\ge i}\sum_{y\le j}A[x,y].
\]
An interval contributes exactly when $i<p_v$ and $q_v<j$. Equation~\eqref{eq:window-degree} therefore gives each candidate degree in constant time.

We fill the tables in increasing order of $j-i$. For $v\in S_{ij}$, we have $i<p_v\le q_v<j$, so both child windows have smaller width, even when $I_v$ is a point interval. Scanning the vertices of a window determines its minimum degree and the two recurrence values in $O(n)$ time. Thus the $O(n^2)$ windows take $O(n^3)$ time in total.

The count preprocessing does not, however, remove the need to compare candidates within each window. Minimizing $d_{G_{ij}}(v)$ is equivalent to maximizing the coupled sum $n_{i,p_v}+n_{q_v,j}$ over $v\in S_{ij}$. Both terms depend on the same candidate, so we cannot maximize them separately. This is structurally analogous to a max-plus product: each pair of window boundaries requires a maximum of sums indexed by a common intermediate choice. Thus, despite the $O(n^2)$ count preprocessing, the $O(n^3)$ transition scan remains the bottleneck of this algorithm.

For the space bound, we store the counts, the two recurrence values, and one attaining first choice for each value in every window. We need the set of tied candidates only while processing the current window, so the total space is $O(n^2)$.

It remains to construct an attaining execution. Use either the maximizing or the minimizing choices, and keep a list of nonempty active windows, initially containing only $G_{0,m+1}$. These windows partition the remaining vertices and have no edges between them. The stored first choice has minimum degree within its window. To obtain a legal choice in the whole remaining graph, we select a stored choice of smallest degree and replace its window by its nonempty children. This choice is legal by Lemma~\ref{lem:interleave}, and the recurrence ensures that it gives the required length.

There are at most $n$ active windows and at most $n$ selections. Computing the degree of each stored choice from~\eqref{eq:window-degree} takes constant time, so reconstruction takes $O(n^2)$ time. This proves Theorem~\ref{thm:A}.
\section{The degree-four bound}\label{sec:degree-four}

We prove that favorable tie-breaking gives a $3/4$ guarantee when the maximum degree is at most four, and that this is tight. We first reduce the lower bound to unit interval graphs.

\subsection{Reduction to unit interval graphs}

\label{lem:simplicial-loss}\label{thm:subcubic}
If $v$ is simplicial, every maximum independent set contains exactly one vertex of the clique $N_G[v]$: it contains at most one by independence and at least one because otherwise $v$ could be added. Replacing that vertex by $v$, and extending any independent set of $G-N_G[v]$ by $v$, gives
\begin{equation}\label{eq:simplicial-loss}
\alpha(G-N_G[v])=\alpha(G)-1.
\end{equation}
Recall from Lemma~\ref{lem:interleave} that component values add. For a nonempty graph, we can also condition on the first choice to obtain
\begin{equation}\label{eq:first-step}
B(G)=1+\max_{d_G(v)=\delta(G)}B(G-N_G[v]).
\end{equation}

\begin{lemma}[Claw elimination]\label{lem:claw-elimination}\label{lem:min-degree-two}
Let $G$ be a nonempty connected interval graph with $\Delta(G)\le4$. If no minimum-degree vertex is simplicial, then $G$ is a unit interval graph.
\end{lemma}

\begin{proof}
We first determine the minimum degree, then exclude induced claws. Vertices of degree zero or one are simplicial, so the hypothesis implies $\delta(G)\ge2$. In an interval representation, let $x$ have minimum right endpoint $r_x$. Every neighbor of $x$ contains $r_x$, so $x$ is simplicial. If $\delta(G)\ge3$, then $d_G(x)\in\{3,4\}$. In the degree-three case, $x$ is a simplicial minimum-degree vertex. In the degree-four case, $N_G[x]$ is a $K_5$ with no outside edges, so connectivity forces $G=K_5$. Both cases contradict the hypothesis. Hence $\delta(G)=2$.

Suppose $G$ contains an induced claw with center $u$ and independent leaves $w_1,w_2,w_3$, ordered so that
$r_{w_1}<\ell_{w_2}\le r_{w_2}<\ell_{w_3}.$
Since $I_u$ meets both outer leaf intervals,
$\ell_u\le r_{w_1}<\ell_{w_2}\le r_{w_2}<\ell_{w_3}\le r_u,$
and hence $I_{w_2}\subseteq I_u$. Every interval meeting $I_{w_2}$ also meets $I_u$, while $w_1,w_3\notin N_G[w_2]$. Thus
$N_G[w_2]\subseteq N_G[u]\setminus\{w_1,w_3\}, \qquad |N_G[w_2]|\le5-2=3.$
As $\delta(G)=2$, vertex $w_2$ has degree two. Its neighbors are $u$ and a vertex adjacent to $u$, so $N_G[w_2]$ is a triangle, contradicting the absence of a simplicial minimum-degree vertex. The graph is therefore claw-free. By Roberts's characterization of indifference graphs as the claw-free interval graphs~\cite{Roberts1969}, it is a unit interval graph. Bogart and West~\cite{BogartWest1999} give a self-contained proof of the equivalence of claw-free interval graphs, proper interval graphs, and unit interval graphs.
\end{proof}

\subsection{A stronger induction hypothesis}

We now consider connected unit interval graphs. The obstacle to a direct induction is that selecting a minimum-degree vertex can reduce the independence number by two. In that case, applying only the bound $B\ge3\alpha/4$ to the remaining graph gives $1+3(\alpha-2)/4$, which falls short of the desired bound by $1/2$. To make up this shortfall, we need to retain some information about the vertices deleted in this step.

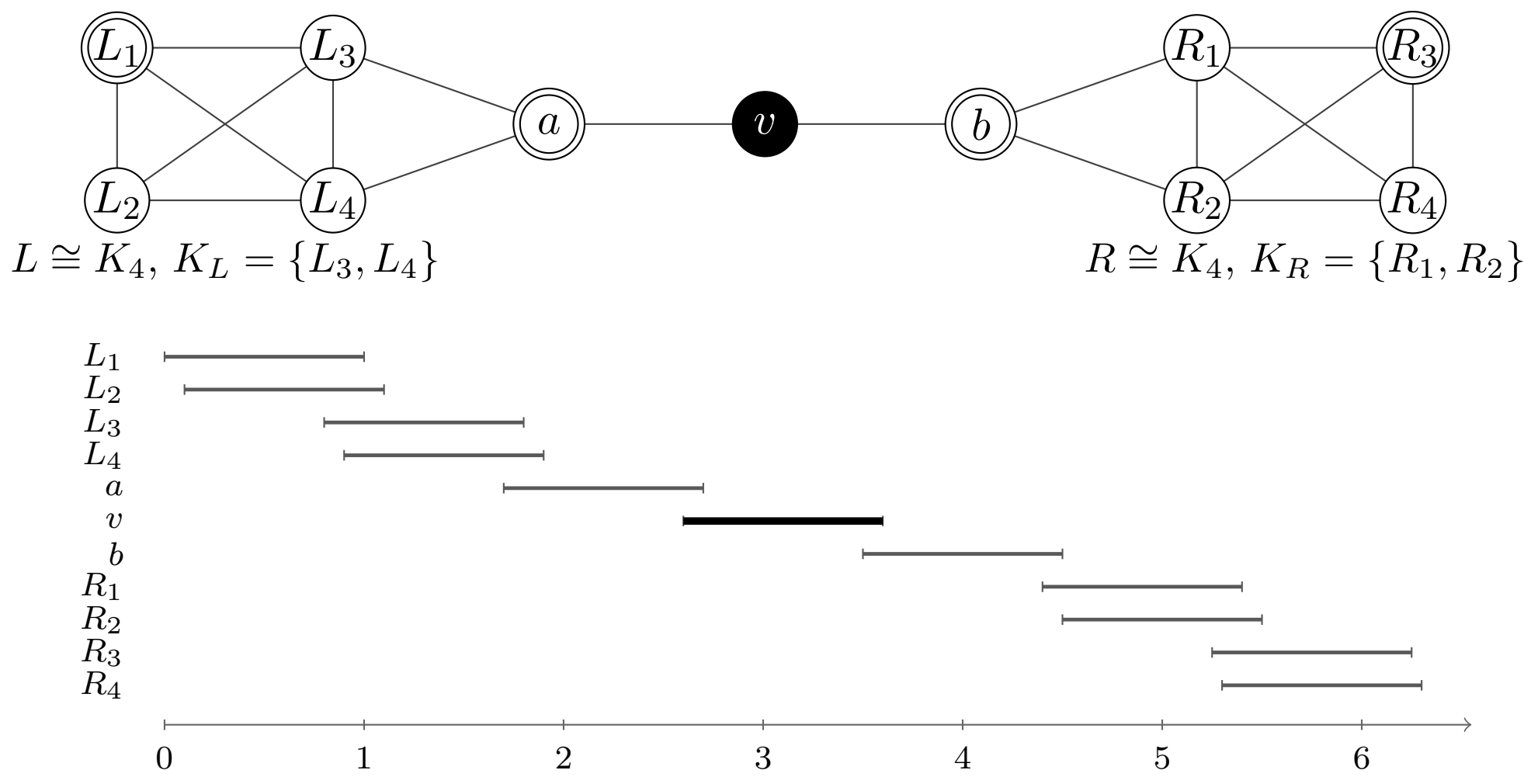
\begin{figure}[htbp]
\centering
\begin{tikzpicture}[x=1.15cm,y=.65cm,
  vertex/.style={circle,draw,fill=white,minimum size=5.5mm,inner sep=0pt,font=\small},
  optimum/.style={vertex,double,double distance=1.2pt}]
\coordinate (p1) at (0,1); \coordinate (p2) at (0,-1);
\coordinate (p3) at (1.6,1); \coordinate (p4) at (1.6,-1);
\coordinate (p5) at (3.2,0); \coordinate (p6) at (4.8,0);
\coordinate (p7) at (6.4,0); \coordinate (p8) at (8,1);
\coordinate (p9) at (8,-1); \coordinate (p10) at (9.6,1);
\coordinate (p11) at (9.6,-1);
\foreach \i/\j in {1/2,1/3,1/4,2/3,2/4,3/4,3/5,4/5,5/6,6/7,7/8,7/9,8/9,8/10,8/11,9/10,9/11,10/11}
  \draw[black!75] (p\i)--(p\j);
\foreach \i/\lab in {2/L_2,3/L_3,4/L_4,8/R_1,9/R_2,11/R_4} \node[vertex] at (p\i) {$\lab$};
\foreach \i/\lab in {1/L_1,5/a,7/b,10/R_3} \node[optimum] at (p\i) {$\lab$};
\node[vertex,fill=black,text=white] at (p6) {$v$};
\node[font=\footnotesize] at (.8,-1.8) {$L\cong K_4$, $K_L=\{L_3,L_4\}$};
\node[font=\footnotesize] at (8.8,-1.8) {$R\cong K_4$, $K_R=\{R_1,R_2\}$};
\end{tikzpicture}

\medskip
\begin{tikzpicture}[x=1.7cm,y=.28cm,font=\scriptsize]
\foreach \i/\s/\lab in {1/0/L_1,2/.1/L_2,3/.8/L_3,4/.9/L_4,5/1.7/a,6/2.6/v,7/3.5/b,8/4.4/R_1,9/4.5/R_2,10/5.25/R_3,11/5.3/R_4}{
  \node[anchor=east] at (-.12,-\i) {$\lab$};
  \draw[black!65,line width=.9pt] (\s,-\i)--({\s+1},-\i);
  \draw[black!65] (\s,{-\i-.16})--(\s,{-\i+.16});
  \draw[black!65] ({\s+1},{-\i-.16})--({\s+1},{-\i+.16});
}
\draw[line width=1.8pt] (2.6,-6)--(3.6,-6);
\draw[->,black!60] (0,-12.2)--(6.55,-12.2);
\foreach \x in {0,1,2,3,4,5,6}{
  \draw[black!60] (\x,-12.04)--(\x,-12.36);
  \node[below] at (\x,-12.36) {$\x$};
}
\end{tikzpicture}

\caption{The graph $J_1$ and a closed unit interval representation, with $B(J_1)=W(J_1)=3<\alpha(J_1)=4$. The black vertex $v$ is the unique minimum-degree vertex; double circles mark a maximum independent set. The boundary pairs are $K_L=\{L_3,L_4\}$ and $K_R=\{R_1,R_2\}$.}
\label{fig:J1}
\end{figure}

The graph $J_1$ in Figure~\ref{fig:J1} illustrates the need for a stronger hypothesis. Its middle vertex $v$ is the unique minimum-degree vertex. Selecting it deletes the path $a-v-b$ and leaves two copies of $K_4$. Each copy still requires a selection, so \MDG\ returns three vertices whereas $\alpha(J_1)=4$. The idea is to keep track of these remaining selections by treating $a$ and $b$ as virtual vertices at the boundaries of the two copies. The general construction of $J_t$ is given later in this section.

A \emph{nonempty boundary state} consists of:
\begin{itemize}
\item a nonempty connected real graph $G$ with an umbrella order;
\item at most one left virtual vertex $z_-$ and at most one right virtual vertex $z_+$;
\item if a left virtual vertex is present, its neighborhood is a clique $K_-$ consisting of the first two or three real vertices;
\item if a right virtual vertex is present, its neighborhood is a clique $K_+$ consisting of the last two or three real vertices.
\end{itemize}
The two boundary cliques may overlap.

Let $\widehat G$ be the graph obtained by adding these vertices and their incident edges to $G$. We require $\Delta(\widehat G)\le4$ and add no edge between the virtual vertices. In the augmented order, $z_-$ comes before all real vertices and $z_+$ comes after them.

Let $k\in\{0,1,2\}$ be the number of virtual vertices, and put
\[
F=\bigcup_{z\text{ virtual}}N_{\widehat G}(z),\qquad A=\alpha(G-F),
\]
with $F=\varnothing$ when $k=0$. The empty state has $G=\varnothing$, $k=0$, $F=\varnothing$, and $A=0$.

The real vertices in $F$ still participate in \MDG. We exclude them only when computing $A=\alpha(G-F)$. Virtual vertices, on the other hand, are never selected, and their incident edges do not contribute to the degrees used by \MDG.

\begin{lemma}\label{lem:augmented-order}
The augmented order is an umbrella order of $\widehat G$.
\end{lemma}

\begin{proof}
We check real and virtual edges separately. No virtual vertex lies between real vertices, so the umbrella condition for a real edge is unchanged. For an edge $z_-x$, every intermediate real vertex lies in the same prefix clique as $x$ and is adjacent to both endpoints. The right-hand case is symmetric. There are no edges between virtual vertices.
\end{proof}

\begin{theorem}\label{thm:boundary}
Every boundary state admits a legal execution on its real graph of length $q$ satisfying

\begin{equation}
4q\ge3A+k. \label{eq:boundary-potential}
\end{equation}
\end{theorem}

The potential $3A+k$ is chosen to account for selections that decrease $A$ by two. Such a selection contributes $4$ to the left-hand side, while the decrease in $A$ costs $6$ on the right. When $A\ge2$ and every minimum-degree choice has this loss, we show that the step creates two new virtual vertices. These supply the missing $2$. On the other hand, if some minimum-degree choice loses at most one, it discards at most one old virtual vertex. Its cost is then at most $3+1=4$.

\subsection{Selections that lose at most one}

For a real vertex $v$, let
\[
H_v=G-N_G[v],\qquad d_F(v)=A-\alpha(H_v-F).
\]
Thus $d_F(v)$ measures the decrease in $\alpha(G-F)$ when $v$ is selected. We first bound this decrease.

\begin{lemma}\label{lem:restricted-loss}
For every maximum independent set $I$ of $G-F$,

\begin{equation}
0\le d_F(v)\le |I\cap N_G[v]|\le2. \label{eq:restricted-loss}
\end{equation}

If $v$ is simplicial, or if $v$ belongs to some maximum independent set of $G-F$, then $d_F(v)\le1$.
\end{lemma}

\begin{proof}
We compare a maximum independent set before and after deletion. Deletion gives $d_F(v)\ge0$. The independent set $I\setminus N_G[v]$ in $H_v-F$ gives $\alpha(H_v-F)\ge A-|I\cap N_G[v]|$. The umbrella-order facts in Section~\ref{sec:preliminaries} give $|I\cap N_G[v]|\le2$. If $v$ is simplicial, its closed neighborhood is a clique and the intersection has size at most one. If $v\in I$, that intersection is $\{v\}$.
\end{proof}

\begin{lemma}\label{lem:low-loss}
Suppose $A\ge2$, and let $v$ be a minimum-degree vertex with $d=d_F(v)\le1$. Put $H=H_v$. Discard every old virtual vertex whose neighborhood meets $N_G[v]$, and retain every other virtual vertex with its entire neighborhood. Then at most one virtual vertex is discarded. Every retained virtual vertex belongs to a unique component of $H$, and each component, with its assigned virtual vertices, is a boundary state. If these states have parameters $(A_i,k_i)$, then

\begin{equation}
\sum_i k_i\ge k-1,
\qquad \sum_i A_i\ge A-d. \label{eq:low-loss-parameters}
\end{equation}
\end{lemma}

\begin{proof}
We first bound the number of discarded virtual vertices, then verify the component states. Suppose, for a contradiction, that both virtual vertices are discarded. Then the consecutive block $N_G[v]$ meets both boundary cliques. If the cliques overlap, their union is $V(G)$, so $G-F$ is empty. If they are disjoint, every vertex outside their union lies between them, since $K_-$ is a prefix and $K_+$ a suffix. Since $N_G[v]$ is consecutive and meets both cliques, it contains all these intermediate vertices. In either case, $V(G)\setminus F\subseteq N_G[v].$ It follows that $\alpha(H-F)=0$ and $d=A\ge2$, a contradiction.

Next, let $K$ be the neighborhood of a retained virtual vertex. Since $N_G[v]$ is a consecutive block disjoint from $K$, the whole of $K$ lies on one side of $N_G[v]$. The clique $K$ remains intact and lies in a unique component of $H$. If $K=K_-$, no real vertex precedes $K$, so it consists of the first $|K|\in\{2,3\}$ vertices of that component. If $K=K_+$, the same argument places it at the last $|K|$ vertices. 

If both boundary cliques are retained, $N_G[v]$ lies strictly between them. No edge joins the surviving prefix to the surviving suffix: such an edge would span $v$ and, by the umbrella property, make its endpoints neighbors of $v$, contrary to their survival. The two retained cliques therefore lie in different components. 

Each augmented component is induced in $\widehat G$ by its real vertices and assigned virtual vertices. Its maximum degree is therefore at most four, and its boundary neighborhoods retain their sizes and positions. 

Finally, let $F'$ be the union of retained neighborhoods in $H$. Assigning each retained virtual vertex once and discarding at most one gives $\sum_i k_i\ge k-1.$ The inclusion $F'\subseteq F\cap V(H)$ and componentwise additivity give $\sum_i A_i=\alpha(H-F')\ge\alpha(H-F)=A-d.$

\end{proof}

\subsection{Selections that lose two}

It remains to consider the case in which every minimum-degree choice loses two. We call a boundary state \emph{difficult} if $A\ge2$ and every minimum-degree vertex $u$ satisfies $d_F(u)=2$.

\begin{lemma}\label{lem:difficult}
In a difficult state, let $v$ be any minimum-degree vertex. Then:

\begin{enumerate}
\item $\delta(G)=2$, and the two neighbors of $v$ can be written $a<v<b$, with $ab\notin E(G)$. The vertices $a,v,b$ are consecutive in the real order.
\item $N_G[v]\cap F=\varnothing$.
\item Every maximum independent set of $G-F$ contains $a,b$ and omits $v$.
\item $3\le d_G(a),d_G(b)\le4$.
\end{enumerate}
\end{lemma}

\begin{proof}
We determine the neighbors of $v$, their relation to $F$, and their degrees. Lemma~\ref{lem:restricted-loss} rules out simplicial minimum-degree vertices, so the minimum-degree argument in Lemma~\ref{lem:claw-elimination} gives $\delta(G)=2$. The two neighbors of $v$ are nonadjacent because $v$ is not simplicial. They must lie on opposite sides of $v$, as same-side neighbors form a clique. Any real vertex between $a$ and $v$, or between $v$ and $b$, would be adjacent to $v$ by the umbrella property, contradicting degree two. 

If $v\in K_-$, then $a<v$ and the prefix property gives $a\in K_-\subseteq F$. If $v\in K_+$, the suffix property similarly gives $b\in K_+\subseteq F$. Thus, if $v\in F$, both $v$ and at least one of $a,b$ belong to $F$. An independent set of $G-F$ then meets $N_G[v]=\{a,v,b\}$ in at most one vertex. This contradicts \eqref{eq:restricted-loss} and $d_F(v)=2$, so $v\notin F$.

If $a\in F$, the intersection can use only the adjacent vertices $v,b$, so it again has size at most one, giving the same contradiction. The case $b\in F$ is symmetric.

Equation~\eqref{eq:restricted-loss} forces every maximum independent set of $G-F$ to meet $N_G[v]$ in exactly two vertices. The only independent pair in the path $a-v-b$ is $\{a,b\}$, so both vertices belong to every such set.

If $d_G(a)=2$, then $a$ is minimum-degree and the difficult-state assumption gives $d_F(a)=2$. But $a$ belongs to a maximum independent set of $G-F$, so Lemma~\ref{lem:restricted-loss} gives $d_F(a)\le1$, a contradiction. Together with $\delta(G)=2$ and $\Delta(G)\le4$, this gives $3\le d_G(a)\le4$. The same argument applies to $b$.
\end{proof}

\begin{lemma}[Boundary splitting]\label{lem:split-boundaries}\label{lem:split-alpha}
In the notation of Lemma~\ref{lem:difficult}, put

$H=G-\{a,v,b\}, \qquad K_L=N_G(a)\setminus\{v\}, \qquad K_R=N_G(b)\setminus\{v\}.$ Then $H$ has exactly two nonempty connected components $L,R$: the prefix strictly before $a$, and the suffix strictly after $b$, respectively. The set $K_L$ consists of the last two or three vertices of $L$ and is a clique; $K_R$ consists of the first two or three vertices of $R$ and is a clique. Regard $a$ as a new right virtual vertex for $L$, and $b$ as a new left virtual vertex for $R$. Retain the old left virtual vertex, if present, in $L$, and the old right virtual vertex, if present, in $R$. Both resulting states are valid, and

\begin{equation}
k_L+k_R=k+2. \label{eq:split-k}
\end{equation}
Writing $F_L,F_R$ for their excluded sets and $A_L=\alpha(L-F_L)$, $A_R=\alpha(R-F_R)$, we also have
\begin{equation}
A_L+A_R=A-2. \label{eq:split-alpha}
\end{equation}
\end{lemma}

\begin{proof}
We verify the component structure and boundary conditions, then prove the equality for $A$. Apart from $v$, the vertex $a$ has no right neighbor. There is no real vertex between $a$ and $v$ or between $v$ and $b$, and $ab$ is not an edge. A neighbor $c>b$ would force $vc$ to be an edge by the umbrella property on $a<v<c$, contradicting $d_G(v)=2$. Thus $K_L$ lies entirely to the left of $a$. Since neighborhoods are consecutive and same-side neighbors form a clique, $K_L$ is a consecutive clique immediately before $a$, of size

$|K_L|=d_G(a)-1\in\{2,3\}.$ The statement for $K_R$ is symmetric.

Deleting the consecutive vertices $a,v,b$ leaves a prefix containing $K_L$ and a suffix containing $K_R$, both nonempty and connected by the umbrella property. An edge joining them would span $a,b$ and force $ab\in E(G)$, so they are exactly the two components.

By Lemma~\ref{lem:difficult}, none of $a,v,b$ belongs to $F$. Since $K_-\subseteq F$ is a prefix of $V(G)$ and $K_+\subseteq F$ is a suffix, each old boundary clique, when present, lies entirely before $a$ or after $b$, respectively, and remains intact in $L$ or $R$.

The augmented left state is induced in $\widehat G$ by $V(L)\cup\{a\}$ and the old left virtual vertex, if present. When we reinterpret $a$ as virtual, its neighborhood is $K_L$. Since $a\notin F$, it is nonadjacent to the old virtual vertex. The boundary conditions and the maximum-degree bound of four are preserved. The same argument verifies the right state. Each side gains one virtual vertex and retains its old one, proving~\eqref{eq:split-k}.

Put $F_H=F\cap V(H)$. The construction gives $F_L\cup F_R=F_H\cup K_L\cup K_R$. Take a maximum independent set $I$ of $G-F$. By Lemma~\ref{lem:difficult}, it contains $a,b$ and omits $v$. It therefore avoids their neighbors $K_L,K_R$. Thus $I\setminus\{a,b\}$ is an independent set of $H$ avoiding $F_L\cup F_R$, with size $A-2$. Additivity over $L,R$ gives $A_L+A_R\ge A-2.$

Conversely, since $F_H\subseteq F_L\cup F_R$, $A_L+A_R =\alpha\bigl(H-(F_L\cup F_R)\bigr) \le\alpha(H-F_H) =A-d_F(v) =A-2.$
\end{proof}

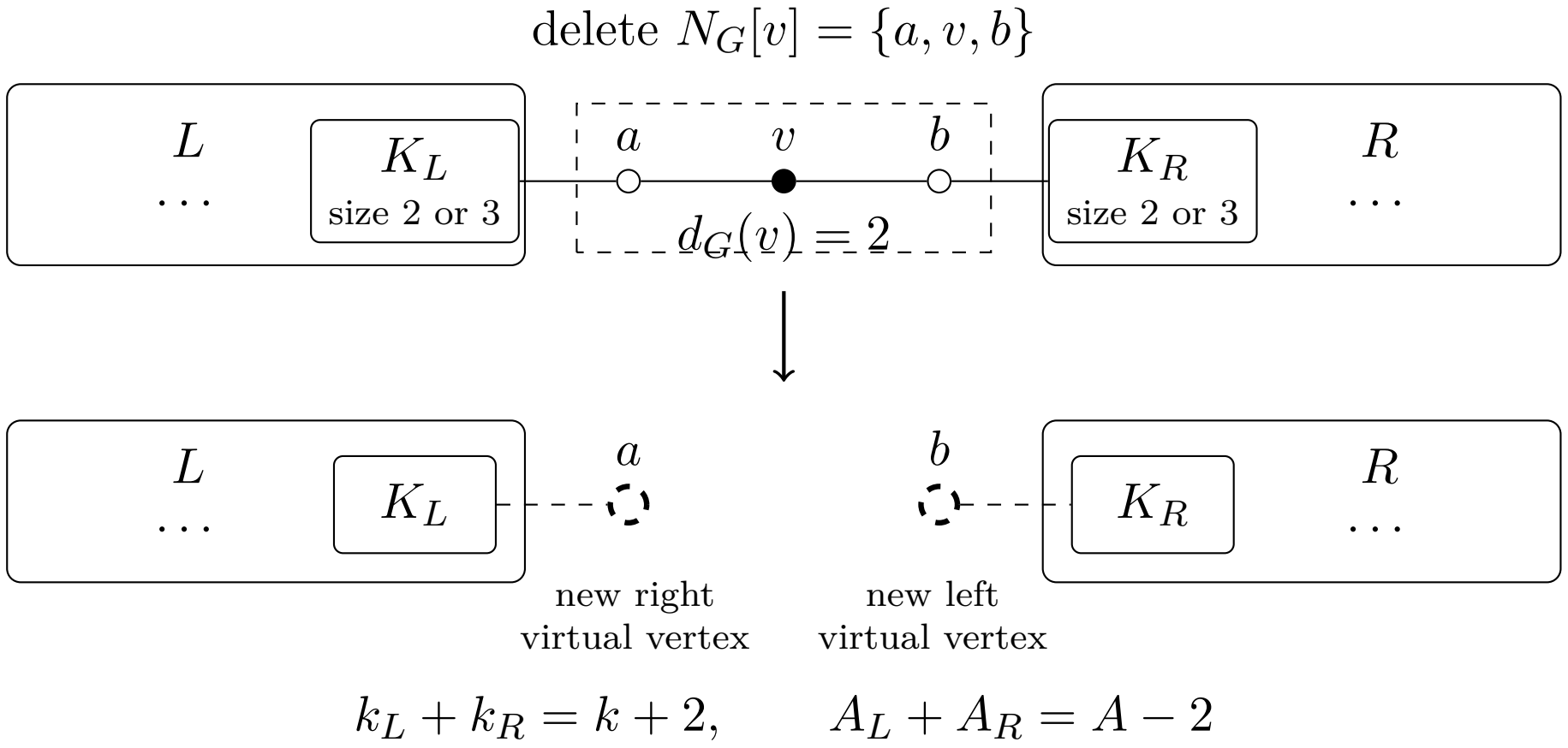
\begin{figure}[tbp]
\centering
\begin{tikzpicture}[x=1cm,y=1cm,font=\small,
  real/.style={circle,draw,fill=white,inner sep=0pt,minimum size=5pt},
  center/.style={real,fill=black},
  virtual/.style={circle,draw,dashed,very thick,fill=white,inner sep=0pt,minimum size=8pt},
  clique/.style={draw,rounded corners=2pt,minimum width=1.25cm,minimum height=.75cm,align=center},
  region/.style={draw,rounded corners=3pt}]
\draw[region] (-6.0,-.65) rectangle (-2.0,.75);
\draw[region] (2.0,-.65) rectangle (6.0,.75);
\node at (-4.6,.32) {$L$}; \node at (4.6,.32) {$R$};
\node at (-4.6,-.18) {$\cdots$}; \node at (4.6,-.18) {$\cdots$};
\node[clique] (kl) at (-2.85,0) {$K_L$\\[-2pt]\scriptsize size $2$ or $3$};
\node[clique] (kr) at (2.85,0) {$K_R$\\[-2pt]\scriptsize size $2$ or $3$};
\node[real,label=above:$a$] (a) at (-1.2,0) {};
\node[center,label=above:$v$,label=below:{$d_G(v)=2$}] (v) at (0,0) {};
\node[real,label=above:$b$] (b) at (1.2,0) {};
\draw (kl)--(a)--(v)--(b)--(kr);
\draw[dashed] (-1.6,-.55) rectangle (1.6,.6);
\node at (0,1.15) {delete $N_G[v]=\{a,v,b\}$};
\draw[->,thick] (0,-.85)--(0,-1.55);
\draw[region] (-6,-3.1) rectangle (-2,-1.85);
\draw[region] (2,-3.1) rectangle (6,-1.85);
\node at (-4.6,-2.2) {$L$}; \node at (4.6,-2.2) {$R$};
\node at (-4.6,-2.7) {$\cdots$}; \node at (4.6,-2.7) {$\cdots$};
\node[clique] (klnew) at (-2.85,-2.5) {$K_L$};
\node[clique] (krnew) at (2.85,-2.5) {$K_R$};
\node[virtual,label=above:$a$] (av) at (-1.2,-2.5) {};
\node[virtual,label=above:$b$] (bv) at (1.2,-2.5) {};
\draw[dashed] (klnew)--(av); \draw[dashed] (bv)--(krnew);
\node[align=center,font=\scriptsize] at (-1.15,-3.35) {new right\\virtual vertex};
\node[align=center,font=\scriptsize] at (1.15,-3.35) {new left\\virtual vertex};
\node at (0,-4.15) {$k_L+k_R=k+2,\qquad A_L+A_R=A-2$};
\end{tikzpicture}
\caption{Splitting a difficult boundary state into $L$ and $R$. Each box represents a clique, and its incident line denotes adjacency to every vertex of that clique. Dashed circles and edges mark the new virtual vertices $a,b$; old virtual vertices are omitted.}
\label{fig:boundary}
\end{figure}

After deleting the real closed neighborhood, we reinterpret $a$ and $b$ as virtual vertices. Each old virtual vertex remains on its original side. Recall that virtual edges serve only to keep track of the boundary. They do not contribute to the degrees used by \MDG.

\subsection{Completing the lower bound}

\begin{proof}[Proof of Theorem~\ref{thm:boundary}]
We use strong induction on the number of real vertices, simultaneously for $k=0,1,2$. The empty state satisfies~\eqref{eq:boundary-potential}. In a nonempty state with $A\le1$, except when $(A,k)=(1,2)$, we have $3A+k\le4$, and any legal execution has length at least one.

The remaining base case is $(A,k)=(1,2)$. To handle it, suppose first that $G=K_n$. Some real vertex is incident with a virtual edge, so $\Delta(\widehat G)\le4$ forces $n\le4$. Since $K_-$ contains at least the first two real vertices and $K_+$ at least the last two, their union covers all $n\le4$ vertices, giving $F=V(G)$ and $A=0$, a contradiction.

Since $G$ is not complete, $\delta(G)\le n-2$, and a legal first selection leaves at least one vertex. Every legal execution therefore has length at least two, giving $4q\ge8\ge5=3A+k.$ 

Now suppose $A\ge2$. We distinguish two cases.

First, suppose a minimum-degree vertex $v$ has $d=d_F(v)\le1$. Select $v$ and form the component states of Lemma~\ref{lem:low-loss}. Each state has fewer real vertices. Induction supplies legal executions of lengths $q_i$ satisfying $4q_i\ge3A_i+k_i.$ By Lemma~\ref{lem:interleave}, interleave these executions and prepend $v$. This gives length $q=1+\sum_iq_i$. By~\eqref{eq:low-loss-parameters},
\[
4q\ge4+3\sum_i A_i+\sum_i k_i\ge4+3(A-d)+k-1\ge3A+k.
\] 

In the second case, every minimum-degree choice loses two, by Lemma~\ref{lem:restricted-loss}. The state is therefore difficult. Choose a minimum-degree vertex $v$ and form the two smaller states of Lemma~\ref{lem:split-boundaries}. Induction supplies legal executions of lengths $q_L,q_R$ satisfying $4q_L\ge3A_L+k_L, \qquad 4q_R\ge3A_R+k_R.$ Interleave these executions and prepend $v$. Equations~\eqref{eq:split-k} and~\eqref{eq:split-alpha} give a legal execution of length $q=1+q_L+q_R$ with
\[
4q=4+4q_L+4q_R\ge4+3(A-2)+k+2=3A+k.
\]

\end{proof}

With the stronger induction complete, we recover the desired bound for unit interval graphs by taking a state with no virtual vertices. The reduction above then extends the bound to all interval graphs.

\begin{theorem}\label{thm:unit-four}\label{thm:interval-four}
Every interval graph $G$ with $\Delta(G)\le4$ satisfies $B(G)\ge3\alpha(G)/4$.
\end{theorem}
\begin{proof}
We apply the boundary theorem to unit interval graphs, then induct for general interval graphs. For a connected unit interval graph, an umbrella order with no virtual vertices is a boundary state with $k=0$, $F=\varnothing$, and $A=\alpha(G)$, so Theorem~\ref{thm:boundary} gives the bound.

For general interval graphs, we use strong induction on $|V(G)|$, starting with the empty graph. If $G$ is disconnected, induction and Lemma~\ref{lem:restriction} give $B(G)=\sum_C B(C)\ge\frac34\sum_C\alpha(C)=\frac34\alpha(G)$, with legal interleaving supplied by Lemma~\ref{lem:interleave}.

If $G$ is connected and has no simplicial minimum-degree vertex, Lemma~\ref{lem:claw-elimination} reduces it to the unit interval case. Otherwise select such a vertex $v$. The smaller interval graph $H=G-N_G[v]$ has maximum degree at most four, so induction and~\eqref{eq:simplicial-loss} give $B(G)\ge1+B(H)\ge1+\frac34\alpha(H)=\frac34\alpha(G)+\frac14$.
\end{proof}

\par\noindent\begin{minipage}{\linewidth}
\subsection{Tightness}

\begin{theorem}\label{thm:J}
For every integer $t\ge1$ there is a connected unit interval graph $J_t$ of maximum degree four such that

\begin{equation}
|V(J_t)|=10t+1,\qquad
\alpha(J_t)=4t,\qquad
B(J_t)=W(J_t)=3t.
\label{eq:J-values}
\end{equation}
\end{theorem}
\end{minipage}\par

\begin{proof}
We give the construction and an interval representation, then determine its independence number and all legal execution lengths. Arrange $t$ disjoint paths $a_j-v_j-b_j$ in a line, with a $K_4$ before the first and a $K_4$ after the last. Between paths $j$ and $j+1$, place a triangle $T_j$ and a $K_4$ called $Q_j$. Add exactly one edge $x_jy_j$ between them, where $x_j\in T_j$ and $y_j\in Q_j$.

Join $b_j$ to the two vertices of $T_j$ other than $x_j$. Join $a_{j+1}$ to exactly two vertices of $Q_j$ other than $y_j$, and let $z_j$ be the fourth vertex of $Q_j$, with no external neighbor. Join $a_1$ to two vertices of the initial $K_4$, and $b_t$ to two vertices of the final $K_4$. Add no other edges.

The graph is connected, and its number of vertices is $3t+7(t-1)+8=10t+1.$ Each path center has degree two, while the path endpoints and triangle vertices have degree three. The $K_4$ vertices have degree three or four. Thus $\Delta(J_t)=4$. Figure~\ref{fig:families} compares this gadget with that of $H_t$.

We next give an interval representation. Use closed intervals of common length three. Set $q_j=11+16(j-1)$ for $1\le j\le t$ and assign the following starts:

\begin{center}
\begin{tabular}{@{}ll@{}}
\toprule
Block & Starts \\
\midrule
Initial $K_4$ & $(0,0,3,3)$ \\
$(a_1,v_1,b_1)$ & $(4,7,8)$ \\
$T_j$, $1\le j<t$ & $(q_j,q_j,q_j+1)$ \\
$Q_j$, $1\le j<t$ & $(q_j+4,q_j+5,q_j+7,q_j+7)$ \\
$(a_{j+1},v_{j+1},b_{j+1})$, $1\le j<t$ & $(q_j+9,q_j+12,q_j+13)$ \\
Final $K_4$ & $(q_t,q_t,q_t+1,q_t+1)$ \\
\bottomrule
\end{tabular}
\end{center}

The starts of $x_j,y_j,z_j$ are $q_j+1,q_j+4,q_j+5$. Only $x_jy_j$ joins $T_j$ to $Q_j$: its start difference is three, and all other differences between these cliques are at least four. The preceding endpoint, at $q_j-3$, meets the first two triangle intervals (difference three) but not $x_j$ (difference four). The following endpoint, at $q_j+9$, meets only the last two $K_4$ intervals.

Within each clique, the starts span at most three. Within each path, consecutive starts differ by three and one, while the endpoints differ by four. Every other pair of blocks has start differences exceeding three. These length-three intervals therefore realize precisely the abstract graph, and scaling gives unit intervals.

\label{sec:j-alpha}

Partition the vertices into $t$ three-vertex paths, $t-1$ triangles, and $t+1$ copies of $K_4$. This gives $\alpha(J_t)\le2t+(t-1)+(t+1)=4t.$ Choose both endpoints of every path, $x_j$ in each $T_j$, and $z_j$ in each $Q_j$. The choices in the two internal blocks are nonadjacent because $x_j$ sees only $y_j$ in the corresponding clique. In each outer $K_4$, choose a vertex not adjacent to its attached path endpoint. All choices are pairwise nonadjacent, giving $\alpha(J_t)=4t$. 

It remains to show that every legal execution selects exactly $3t$ vertices. While a path center remains, the following properties hold:

\begin{enumerate}
\item Every surviving path is intact, with center degree two and endpoint degrees three.
\item Every $K_4$ is intact, so every one of its vertices has degree at least three.
\item Each surviving triangle is intact. While its preceding path survives, all its vertices have degree three. After that path is deleted, its two nonconnector vertices have degree two and $x_j$ has degree three.
\item A deleted triangle has been removed by selecting one of its two exposed nonconnector vertices. Such a deletion removes only the triangle; it decreases the degree of $y_j$ from four to three and affects no surviving path.

\end{enumerate}
These properties hold initially. To see that they persist, suppose a center remains. The minimum degree is then two, and the eligible vertices are exactly the surviving centers and exposed nonconnector triangle vertices. Selecting a center deletes its path, possibly exposing the following triangle and removing an attachment from the preceding $K_4$. Every $K_4$ vertex still has its three neighbors within the clique. Selecting an exposed nonconnector deletes its triangle without deleting $y_j$ or a surviving path endpoint. Both operations preserve the invariant.

Thus every path is removed by selecting its center, and no $K_4$ vertex is selected while a center remains. Once all centers are gone, each surviving triangle still has two degree-two vertices, whereas all $K_4$ degrees are at least three. Thus the remaining triangles must each be removed by one selection before any $K_4$ vertex can be selected. The residual graph then consists of $t+1$ disjoint intact $K_4$s. Every legal execution has length $t+(t-1)+(t+1)=3t.$

\end{proof}

It is helpful to return to $J_1$ and see where equality holds in the stronger bound. The initial state has $(A,k)=(4,0)$. After selecting $v$, each remaining $K_4$ has one virtual boundary adjacent to two of its vertices, so $(A_i,k_i)=(1,1)$. Each is a base case requiring one selection, and $4q_i=3A_i+k_i=4$. At the root, $q=3$ and $4q=3A+k=12$.

Combining Theorem~\ref{thm:interval-four} with the family $J_t$ gives $c_4=3/4$, also within unit interval graphs, and completes Theorem~\ref{thm:B}.

\section{Degree bounds of at least five}\label{sec:degree-five}

We now show that the guarantee drops to $2/3$ when the maximum-degree bound is at least five. Since the lower bound is known, it remains to construct a family whose ratio approaches it even under favorable tie-breaking.

\subsection{The lower bound}
\label{lem:two-loss-split}\label{lem:two-thirds}
Chaplick et al.~\cite[Lemma~5 and Theorem~2]{ChaplickEtAl2024} proved that every legal execution on an interval graph has length at least $2\alpha(G)/3$. Their component-counting argument tells us a little more: this ratio is an infimum, and no finite nonempty graph attains it.

Fix a maximum independent set, and let $m_j$ count the steps that delete $j$ of its surviving vertices. Each step has $j\le2$, and a step with $j=2$ increases the number of components by at least one. If $c$ is the initial number of components, the argument gives $m_0+m_1\ge c+m_2$. Since $\alpha(G)=m_1+2m_2$ and the execution length is $q=m_0+m_1+m_2$, we obtain
\[
3q-2\alpha(G)=2m_0+(m_0+m_1-m_2)\ge c\ge1.
\]
Thus every execution on a finite nonempty interval graph has ratio strictly greater than $2/3$.

\subsection{Tightness}

\begin{theorem}\label{thm:H}
For every integer $t\ge2$ there is a connected unit interval graph $H_t$ of maximum degree five such that

\begin{equation}
|V(H_t)|=8t+3,\qquad
\alpha(H_t)=3t+1,\qquad
B(H_t)=W(H_t)=2t+1.
\label{eq:H-values}
\end{equation}

The same construction at $t=1$ gives $H_1\cong J_1$, an eleven-vertex graph of maximum degree four.
\end{theorem}

\begin{figure}[H]
\centering
\begin{tikzpicture}[x=1cm,y=1cm,font=\small,
  vertex/.style={circle,draw,fill=white,inner sep=0pt,minimum size=5pt},
  selected/.style={circle,draw,double,double distance=1.0pt,fill=white,inner sep=0pt,minimum size=6pt},
  center/.style={vertex,fill=black,minimum size=6pt},
  lab/.style={font=\scriptsize}]
\node[anchor=west] at (-6.25,2.1) {(a) $J_t$: internal triangle--$K_4$ gadget, $\Delta=4$};
\node[selected,label=above:$a_j$] (ja) at (-5.8,0) {};
\node[center,label=below:$v_j$] (jv) at (-4.65,0) {};
\node[selected,label=below:$b_j$] (jb) at (-3.5,0) {};
\node[vertex] (ju) at (-2.25,.7) {};
\node[vertex] (jw) at (-2.25,-.7) {};
\node[selected,label=above:$x_j$] (jx) at (-1.1,0) {};
\node[vertex,label=above:$y_j$] (jy) at (.25,0) {};
\node[selected,label=above:$z_j$] (jz) at (1.2,.85) {};
\node[vertex] (jq) at (2.15,0) {};
\node[vertex] (jr) at (1.2,-.85) {};
\node[selected,label=below:$a_{j+1}$] (jaa) at (3.5,0) {};
\node[center,label=below:$v_{j+1}$] (jvv) at (4.65,0) {};
\node[selected,label=above:$b_{j+1}$] (jbb) at (5.8,0) {};
\draw (ja)--(jv)--(jb) (jb)--(ju)--(jw)--(jx)--(ju) (jb)--(jw);
\draw (jx)--(jy);
\foreach \u/\v in {jy/jz,jy/jq,jy/jr,jz/jq,jz/jr,jq/jr} \draw (\u)--(\v);
\draw (jq)--(jaa) (jr)--(jaa) (jaa)--(jvv)--(jbb);
\draw[densely dotted] (-6.2,0)--(ja) (jbb)--(6.2,0);
\node at (-1.95,1.35) {$T_j$}; \node at (2.0,1.5) {$Q_j\cong K_4$};
\node[lab,align=center] at (-2.0,-1.35) {two nonconnector vertices\\can become eligible};
\node[lab] at (1.8,-1.4) {$z_j$: no external neighbor};

\node[anchor=west] at (-6.25,-2.3) {(b) $H_t$: internal $K_5$ gadget, $\Delta=5$};
\node[selected,label=above:$a_j$] (ha) at (-5.8,-4.1) {};
\node[center,label=below:$v_j$] (hv) at (-4.65,-4.1) {};
\node[selected,label=below:$b_j$] (hb) at (-3.5,-4.1) {};
\node[vertex] (hu) at (-1.3,-3.35) {};
\node[vertex] (hw) at (-1.3,-4.85) {};
\node[selected,label=below:{$\scriptstyle\text{middle}$}] (hm) at (0,-5.15) {};
\node[vertex] (hq) at (1.3,-4.85) {};
\node[vertex] (hr) at (1.3,-3.35) {};
\node[selected,label=below:$a_{j+1}$] (haa) at (3.5,-4.1) {};
\node[center,label=below:$v_{j+1}$] (hvv) at (4.65,-4.1) {};
\node[selected,label=above:$b_{j+1}$] (hbb) at (5.8,-4.1) {};
\foreach \u/\v in {hu/hw,hu/hm,hu/hq,hu/hr,hw/hm,hw/hq,hw/hr,hm/hq,hm/hr,hq/hr} \draw (\u)--(\v);
\draw (ha)--(hv)--(hb) (hb)--(hu) (hb)--(hw);
\draw (hq)--(haa) (hr)--(haa) (haa)--(hvv)--(hbb);
\draw[densely dotted] (-6.2,-4.1)--(ha) (hbb)--(6.2,-4.1);
\node at (0,-2.95) {$C_j\cong K_5$};
\node[lab] at (0,-5.9) {the middle vertex has no external neighbor};
\node[center] at (-4.7,-6.65) {};
\node[anchor=west,lab] at (-4.45,-6.65) {degree-two path center};
\node[selected] at (.15,-6.65) {};
\node[anchor=west,lab] at (.4,-6.65) {vertex of the exhibited maximum independent set};
\end{tikzpicture}
\caption{Repeating gadgets of $J_t$ and $H_t$ for $1\le j<t$; outer $K_4$ blocks are omitted. Double circles mark vertices of the exhibited maximum independent sets, and black vertices mark path centers selected in every legal execution. Edge crossings inside cliques do not represent vertices.}
\label{fig:families}
\end{figure}
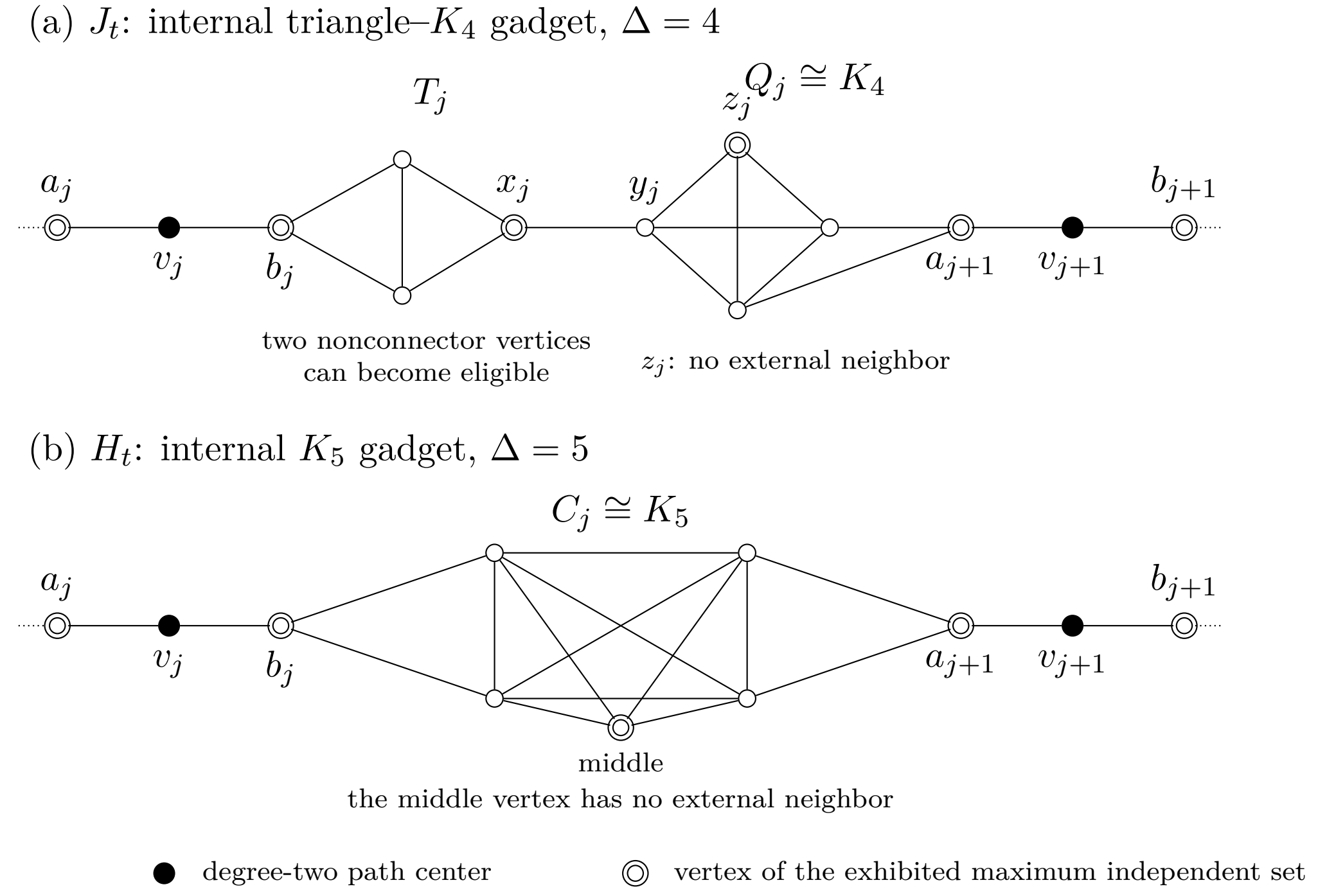

The two families differ in the order of legal selections. In $H_t$, every center is selected before any clique vertex. In $J_t$, exposed nonconnector triangle vertices may be selected between centers. Double circles in the figure mark independent-set vertices rather than choices of \MDG.

\begin{proof}

We verify the construction and its interval representation, then compute the independence number and execution lengths. Consider Figure~\ref{fig:families}(b). Take disjoint cliques $C_0,\dots,C_t$, with end cliques $C_0,C_t$ of size four and internal cliques of size five. Between $C_{j-1}$ and $C_j$, insert a disjoint three-vertex path $a_j-v_j-b_j$. Join $a_j$ to the last two vertices of $C_{j-1}$ and $b_j$ to the first two vertices of $C_j$. In an internal $K_5$, the first and last pairs are disjoint, leaving a middle vertex with no external neighbor. Add no other edges.

There are $8+5(t-1)+3t=8t+3$ vertices. Path centers have degree two and path endpoints have degree three. Each internal clique vertex has four neighbors within its clique and at most one outside it. End-clique vertices have degree three or four. The graph is connected and has the claimed maximum degree. 

To verify that $H_t$ is a unit interval graph, use closed intervals of length three, with equal starts representing distinct vertices. Put $q_j=11+12(j-1),\qquad 1\le j\le t.$ Assign starts as follows:

\begin{center}
\begin{tabular}{@{}ll@{}}
\toprule
Block & Starts \\
\midrule
$C_0$ & $(0,0,3,3)$ \\
$(a_1,v_1,b_1)$ & $(4,7,8)$ \\
$C_j$, $1\le j<t$ & $(q_j,q_j,q_j+1,q_j+3,q_j+3)$ \\
$(a_j,v_j,b_j)$, $2\le j\le t$ & $(q_{j-1}+5,q_{j-1}+8,q_{j-1}+9)$ \\
$C_t$ & $(q_t,q_t,q_t+1,q_t+1)$ \\
\bottomrule
\end{tabular}
\end{center}

Each clique has start span at most three. The preceding path endpoint of $C_j$, at $q_j-3$, meets exactly its first two intervals. For an internal clique, the following endpoint, at $q_j+5$, meets exactly its last two intervals. Consecutive path starts differ by three and one, whereas the endpoints differ by four and are nonadjacent. All remaining cross-block differences exceed three, so the coordinates realize the abstract graph. Scaling by $1/3$ gives unit intervals.

Partition the vertices into the $t+1$ cliques and $t$ three-vertex paths. An independent set takes at most one vertex from each clique and two from each path. Hence $\alpha(H_t)\le3t+1$. Choose both endpoints of each path, the middle vertex of each internal $K_5$, a vertex from the first two of $C_0$, and a vertex from the last two of $C_t$. These choices avoid all attachment edges and are pairwise nonadjacent. Hence $\alpha(H_t)=3t+1$.

Finally, consider an arbitrary legal execution of \MDG. Initially, the centers are exactly the degree-two vertices. Selecting one deletes its path while leaving all cliques intact and all other paths unchanged. The surviving clique vertices and path endpoints retain degree at least three. Thus, as long as a center survives, the algorithm must select a center.

After the $t$ centers have been selected, the residual graph is the disjoint union of $t+1$ intact cliques. A legal execution selects one vertex per clique, for a total of $t+(t+1)=2t+1$. \end{proof}

In particular,

\begin{equation}
\frac{B(H_t)}{\alpha(H_t)}
=\frac{2t+1}{3t+1}
=\frac23+\frac{1}{9t+3}\longrightarrow\frac23.
\label{eq:H-limit}
\end{equation}

For $t=2$, this is a 19-vertex graph with $B=5$ and $\alpha=7$, so $4B<3\alpha$.

The known lower bound gives $c_d\ge2/3$. For every fixed $d\ge5$, the graphs $H_t$, $t\ge2$, belong to the defining class, and their ratios tend to $2/3$ by~\eqref{eq:H-limit}. Hence $c_d=2/3$, proving Theorem~\ref{thm:C}. The same argument applies to unit interval graphs and to the class with no degree bound.

\section{Conclusion and open questions}
\label{sec:conclusion}

We conclude with two open questions. The first is the complexity of computing $B(G)$ and $W(G)$ on chordal graphs. The question for $B(G)$ was posed by Chaplick et al.~\cite[Section~4]{ChaplickEtAl2024}.

In a clique-tree representation, the maximal cliques containing a vertex form a connected subtree. Deleting $N_G[v]$ removes the vertices whose subtrees meet that of $v$, potentially leaving several pendant subtrees rather than the two linearly ordered pieces of an interval model. Executions on the resulting components can still be interleaved by Lemma~\ref{lem:interleave}. The difficulty lies in describing the recursive subproblems and their degree counts with polynomially many states. In this setting, two endpoint indices no longer describe a branching boundary.

A second question concerns the size of the equality graphs for degree four. Our family $J_t$ satisfies $\alpha(J_t)=4t$, $B(J_t)=3t$, and has $10t+1$ vertices. Must every connected unit interval graph $G$ with $\Delta(G)\le4$, $\alpha(G)=4t$, and $B(G)=3t$ have at least $10t+1$ vertices?

\section*{Funding}
This research did not receive any specific grant from funding agencies in the public, commercial, or not-for-profit sectors.

\section*{Declaration of competing interests}
The author declares that there are no competing interests.

\section*{Data availability}
No datasets were generated or analysed for this study.

\section*{Declaration of generative AI and AI-assisted technologies in the manuscript preparation process}
During the preparation of this work, the author used OpenAI ChatGPT to improve the readability and language of the manuscript, and takes full responsibility for the content of the published article.

\end{document}